\documentclass[runningheads,orivec,envcountsame]{llncs}
\usepackage{amsmath,amssymb}
\usepackage{xspace}
\usepackage{cite}
\usepackage{hyperref}
\usepackage[disable]{todonotes}
\usepackage{color}
\usepackage{graphicx}
\newcommand{\TWO}{\ensuremath{\{\textup{\texttt{0}},\textup{\texttt{1}}\}}}
\newcommand{\Lrm}[1]{\ensuremath{\mathrm{L}^{#1}}}
\newcommand{\ELL}[1]{\ensuremath{\mathrm{L}^{#1}}}

\newcommand{\ELLone}{\ELL{1}}
\newcommand{\ELLtwo}{\ELL{2}}

\newcommand{\SOB}[1]{\ensuremath{\mathrm{W}^{#1}}}
\newcommand{\IN}{\mathbb{N}}
\newcommand{\IZ}{\mathbb{Z}}
\newcommand{\IR}{\mathbb{R}}
\newcommand{\IC}{\mathbb{C}}
\newcommand{\calA}{\mathcal{A}}
\newcommand{\calB}{\mathcal{B}}
\newcommand{\calC}{\mathcal{C}}
\newcommand{\calD}{\mathcal{D}}
\newcommand{\calF}{\mathcal{F}}
\newcommand{\calO}{\mathcal{O}}
\newcommand{\calS}{\mathcal{S}}
\newcommand{\Entropy}{\eta}
\newcommand{\Martin}[1]{\textcolor{red}{#1}} 
\newcommand{\Aras}[1]{\textcolor{blue}{#1}} 
\newcommand{\IS}{\mathbb{S}}
\newcommand{\IT}{\mathbb{T}}
\newcommand{\Log}{\operatorname{Log}}
\newcommand{\Bochner}{\mathcal{B}}
\newcommand{\sharpP}{\text{\#\sf P}\xspace}
\newcommand{\classFP}{\text{\sf FP}\xspace}
\newcommand{\classFPSPACE}{\text{\sf FPSPACE}\xspace}
\newcommand{\poly}{\operatorname{poly}}
\spnewtheorem{fact}[theorem]{Fact}{\bfseries}{\itshape}

\title{Second-Order Parameterizations 
for the Complexity Theory of Integrable Functions}
\titlerunning{Second-Order Parameterizations of Spaces of Integrable Functions}
\author{%
Aras Bacho\inst{1}\orcidID{0000-0002-2333-0884}
\and
Martin Ziegler\inst{2}\orcidID{0000-0001-6734-7875}}

\institute{%
California Institute of Technology, USA (Caltech) \\
\email{bacho@caltech.edu}, 
\url{https://arasbacho.github.io} 
\linebreak \and 
Korea Advanced Institute of Science and Technology (KAIST) \\
\email{ziegler@kaist.ac.kr}, 
\url{http://ziegler.theoryofcomputation.asia}
}
\authorrunning{Aras Bacho (Caltech) \and Martin Ziegler (KAIST)}
\date{}

\begin{document}
\makeatletter
\renewcommand\maketitle{\newpage
  \refstepcounter{chapter}%
  \stepcounter{section}%
  \setcounter{section}{0}%
  \setcounter{subsection}{0}%
  \setcounter{figure}{0}
  \setcounter{table}{0}
  \setcounter{equation}{0}
  \setcounter{footnote}{0}%
  \begingroup
    \parindent=\z@
    \renewcommand\thefootnote{\@fnsymbol\c@footnote}%
    \if@twocolumn
      \ifnum \col@number=\@ne
        \@maketitle
      \else
        \twocolumn[\@maketitle]%
      \fi
    \else
      \newpage
      \global\@topnum\z@   
      \@maketitle
    \fi
    \thispagestyle{empty}\@thanks
    \def\\{\unskip\ \ignorespaces}\def\inst##1{\unskip{}}%
    \def\thanks##1{\unskip{}}\def\fnmsep{\unskip}%
    \instindent=\hsize
    \advance\instindent by-\headlineindent
    \if@runhead
       \if!\the\titlerunning!\else
         \edef\@title{\the\titlerunning}%
       \fi
       \global\setbox\titrun=\hbox{\small\rm\unboldmath\ignorespaces\@title}%
       \ifdim\wd\titrun>\instindent
          \typeout{Title too long for running head. Please supply}%
          \typeout{a shorter form with \string\titlerunning\space prior to
                   \string\maketitle}%
          \global\setbox\titrun=\hbox{\small\rm
          Title Suppressed Due to Excessive Length}%
       \fi
       \xdef\@title{\copy\titrun}%
    \fi
    \if!\the\tocauthor!\relax
      {\def\and{\noexpand\protect\noexpand\and}%
      \protected@xdef\toc@uthor{\@author}}%
    \else
      \def\\{\noexpand\protect\noexpand\newline}%
      \protected@xdef\scratch{\the\tocauthor}%
      \protected@xdef\toc@uthor{\scratch}%
    \fi
    \if@runhead
       \if!\the\authorrunning!
         \value{@inst}=\value{@auth}%
         \setcounter{@auth}{1}%
       \else
         \edef\@author{\the\authorrunning}%
       \fi
       \global\setbox\authrun=\hbox{\small\unboldmath\@author\unskip}%
       \ifdim\wd\authrun>\instindent
          \typeout{Names of authors too long for running head. Please supply}%
          \typeout{a shorter form with \string\authorrunning\space prior to
                   \string\maketitle}%
          \global\setbox\authrun=\hbox{\small\rm
          Authors Suppressed Due to Excessive Length}%
       \fi
       \xdef\@author{\copy\authrun}%
       \markboth{\@author}{\@title}%
     \fi
  \endgroup
  \setcounter{footnote}{\fnnstart}%
  \clearheadinfo}

\def\institutename{\par
 \begingroup
 \parskip=\z@
 \parindent=\z@
 \setcounter{@inst}{1}%
 \def\and{\qquad\stepcounter{@inst}%
 \noindent$^{\the@inst}$\enspace\ignorespaces}%
 \setbox0=\vbox{\def\thanks##1{}\@institute}%
 \ifnum\c@@inst=1\relax
   \gdef\fnnstart{0}%
 \else
   \xdef\fnnstart{\c@@inst}%
   \setcounter{@inst}{1}%
   \noindent$^{\the@inst}$\enspace
 \fi
 \ignorespaces
 \@institute\par
 \endgroup}

\makeatother

\maketitle

\begin{abstract}
We develop a unified second-order parameterized complexity theory for spaces of integrable functions.
This generalizes the well-established case of
continuous functions.
Specifically we prove the mutual linear equivalence of three
natural parameterizations of the space $\Lrm{p}$ of $p$-integrable complex functions on the real unit interval:
(binary) $\Lrm{p}$-modulus, rate of convergence of Fourier series, 
and rate of approximation by step functions.
\end{abstract}

\renewcommand{\contentsname}{}
\setcounter{tocdepth}{3}
\begin{minipage}{0.9\textwidth}
\small
\tableofcontents
\end{minipage}
\setcounter{secnumdepth}{3}

\renewcommand{\Martin}[1]{} \renewcommand{\Aras}[1]{}  

\section{Introduction}
Computability and complexity investigations regarding real or complex functions
provide a rigorous foundation to numerical computations \cite{Wei00,BC06}.
Such investigations have exhibited surprising connections 
to discrete computability \cite{Spe59,PR79}
and to famous discrete complexity conjectures
\cite{Fri84,Ko91,Kaw10,KSZ17,KPSZ23}.

Note that the latter works investigate computational complexity in the
Banach space $\calC(X)$ of continuous, or $\calC^k(X)$ of continuously differentiable, 
functions over some compact Euclidean domain $X$.
However such spaces are in many situations known unsuitable for the mathematical solution theory of partial differential equations (PDEs): Heat and Wave and 
Schr\"{o}dinger and Navier-Stokes Equation naturally `live'
in different Banach spaces, such as $\ELL{p}(X)$ of $p$-integrable functions over $X$.
Accordingly also qualitative computability investigations of PDEs
make more sense in such advanced function spaces from Analysis
\cite{PR81,WZ02a,DBLP:conf/birthday/Sun0020}.
Underlying these works, and key to the well-established general definition and theory 
of computable Banach spaces \cite{PR89,Kun04a,DBLP:journals/mlq/BrattkaD07}, is the observation that 
various natural encodings \cite{KW85,Sch06}---based on different mathematically common choices of countable dense subsets
(including rational polynomials, Fourier series, and step functions)---turn out to be computably equivalent: 
qualitatively. 

\medskip
\emph{Quantitative} complexity investigations on the other hand 
of, say, PDEs have so far been hampered by 
the lack of a (well-justified and generally accepted) 
\emph{definition} of computational complexity for $\ELL{p}(X)$.
Previous work \cite{LimZiegler25} has refined the aforementioned generic coding theory
from qualitative to quantitative: for \emph{compact} metric spaces.
However $\ELL{p}(X)$ is not compact, not even $\sigma$-compact for infinite compact $X$. 
Neither is $\calC(X)$,
but the latter space canonically comes with
a `natural' parameterization in terms of \emph{modulus of continuity};
and that is commonly used to define and investigate parameterized computational complexity,
based on its induced monotone covering by compact subsets of $\calC(X)$.
Integrable functions on the other hand 
need not be continuous and thus cannot 
be parameterized in this way. 
This raises the following question \cite{Hilbert}:

\begin{question}
\label{q:Main}
Fix some compact Euclidean domain $X$.
How parameterize the space 
$\ELL{p}(X)$ of $p$-integrable functions
as monotone cover of compact subsets?
\end{question}
Previous work \cite[\S3.1+\S5.3]{Ste17} has proposed 
a natural such parameterization of $\ELL{p}(X)$. 
We formalize two further, and equally natural,
parameterizations of $\ELL{p}(X)$. 
And we establish all three to be linearly equivalent, that is,
mutually bounded up to constant factors in both argument and value.
In particular,
from a linear (or polynomial) perspective,
the three parameterizations induce the same notion of computational complexity.
This arguably answers Question~\ref{q:Main}.

\subsection{Overview}
\label{s:Overview}

Subsection~\ref{ss:Parameterizations} recalls and emphasizes 
the ubiquity of (often implicitly) parameterizing
a computational problem, and the various impact 
(linear, polynomial, exponential) of different
parameterizations on the induced computational complexity.
Subsection~\ref{ss:Continuous} reviews well-established
parameterizations of the particular 
space of continuous functions on a compact domain, 
together with their justifications.

In Section~\ref{s:Integrable} we adapt and generalize these considerations
to the Banach space $\ELL{p}$ of $p$-integrable functions:
Subsection~\ref{ss:LpModulus} recalls the $\ELL{p}$-modulus
proposed in \cite[\S3.1+\S5.3]{Ste17} for parameterizing $\ELL{p}$.
Subsections~\ref{ss:StepModulus} and \ref{ss:FourierRate}
introduce two further parameterizations: 
quantitative rates of approximation by step functions and by Fourier series, respectively.
And Subsection~\ref{ss:Equivalent} establishes all three
as mutually bounded up to constant factors in arguments and values.
Thus, for complexity investigations
on spaces of integrable functions, these three natural parameterizations
are all suitable and linearly equivalent.
Proofs are deferred to Subsection~\ref{ss:Proof}.

Section~\ref{s:Perspectives} collects some extensions and examples:
Subsection~\ref{ss:Examples} constructs functions with asymptotically 
(small and with) large moduli/rates. 
Subsection~\ref{ss:Banach} generalizes Step and Fourier rate 
from $\ELL{p}(\IT)$ and $\ELL{p}(\IS)$
to Sobolev spaces over $\IT/\IS$;
and further on to any separable Banach space
with respect to some given Schauder system.
Subsection~\ref{ss:Dimension} discusses 
integrable functions over Euclidean domains
of dimensions larger than one.

\subsection{Parameterizing Problems}
\label{ss:Parameterizations}
This subsection recalls the ubiquity, and---often implicit---choices, 
for \emph{parameterizing} both discrete (Subsubsection~\ref{sss:Discrete})
and continuous (Subsubsections~\ref{sss:Compact}+ \ref{sss:SigmaCompact})
computational problems.
It emphasizes the impact of different parameterizations 
on the induced computational complexity.

\subsubsection{Classical complexity theory}
\label{sss:Discrete}
focuses on functions or decision problems over discrete domains.  
Any algorithm $\calA$ computing a total discrete function
$f : \TWO^* \to \TWO^*$
admits a worst-case \emph{time complexity} bound $T : \IN \to \IN$ depending only on the input length $n$:
Consider the (by totality hypothesis finite) number $T_\calA(\vec x)\in\IN$ of steps $\calA$ makes
on input $\vec x\in\TWO^*$, and take its maximum over all (again finitely many) inputs of length $|\vec x|\leq n$.
This folklore observation, that worst-case time complexity is well-defined,
extends seamlessly to \emph{space}/memory complexity $S(n)$, counting the number of tape cells accessed.
Particularly desirable is of course polynomial runtime $T(n)\leq n^{\calO(1)}$, captured by the complexity class $\classFP$;
or at least polynomial memory, captured by $\classFPSPACE$.

Note that this observation applies also for unary (instead of binary) input encoding,
but with a tremendous change in the induced complexity. 
Integer factorization for example is believed \emph{in}feasible in polynomial time---as underlying the security of the RSA cryptosystem---with 
respect to binary input length $n\gtrsim1000$ (corresponding to integer keys $N$ of $\gtrsim1000$ bits);
while \emph{Sieve of Eratosthenes} can find integer factors even in linear time, 
with respect to the unary/value parameter $N\approx2^{1000}$. 

On the other hand, proceeding from binary to decimal encoding or vice versa
changes the input length only by a constant factor;
hence computational cost estimates of $T\big(\calO(n)\big)$,
as opposed to $T(n)$, remain `robust' under such changes.
Proceeding from a multitape to a one-tape Turing machine is known
to incur an increase in running time increase most polynomial,
meaning that $\poly(T)$ is robust with respect to the latter;
and $\poly\big(T(\calO(n))\big)$ under both.

Our main results in Subsection~\ref{ss:Equivalent}
establish similar robustness with respect to various natural
but superficially unrelated ways of parameterizing 
spaces of integrable functions.

\subsubsection{Complexity on Compact Metric Domains:}
\label{sss:Compact}
Many real-world problems require computations on \emph{continuous} data.  
A prime setting here is \emph{Computable Analysis} \cite[\S3]{Wei00}, which considers inputs and outputs given by infinite sequences of approximations.
Computational cost thus is measured in dependence not on the (infinite) input length
but on the output approximation error $2^{-n}$, 
commonly parameterized also by the natural number $n$; cmp. \cite[\S7.1]{Wei00} and \cite[\S2.4]{Ko91}.

The folklore observation from Subsubsection~\ref{sss:Discrete} adapts from the discrete to this continuous setting, under suitable hypotheses.
For instance, if a function $g : X \subseteq \IR \to \IC$ is total on \emph{compact} domain $X$, 
then any correct\footnote{with respect to an \emph{admissible} encoding of $X$ and $\IC$ \cite{KW85,Sch06,LimZiegler25}} 
computational procedure for $g$ will have a worst-case time complexity $t(n)$ 
that depends only on $n$; cmp. \cite[Theorem~7.2.7]{Wei00}.
Informally, no matter which input $x\in X$,
the computational cost to guarantee $n$ digits (bits) of correct output can be bounded uniformly by $t(n)$.
Again, this observation applies also to the `unary' error bound $1/n$ (instead of `binary' $2^{-n}$),
but with a possibly tremendous change in the induced time complexity $t$.

For example, consider the definite Riemann integral  
of some function $g:[0;1]\to[0;1]$ promised to be 
non-expansive aka 1-Lipschitz:
Approximating $\int_0^1 g(t)\,dt$ up to error $1/n$ takes
$\calO(n)$ samples $g(x)$, but approximating up to $2^{-n}$ takes $\calO(2^n)$ samples.
Note that approximation up to $1/n$ can be calculated in time polynomial in $n$;
and approximation up to error $2^{-n}$ can using memory of size polynomial in $n$,
whereas polynomial runtime there is infeasible---even for \emph{fixed} smooth 
polynomial-time computable $g$, unless $\sharpP_1=\classFP_1$ \cite[Theorem~5.32]{Ko91}.

\begin{example}
\label{x:Modulus}
An information-theoretic lower bound on the worst-case time complexity of any function $g:X\subseteq\IR\to\IC$
arises from the `sensitivity' of its values to perturbations of the argument
at smaller and smaller scales,
captured by $g$'s (binary) \emph{modulus} of uniform continuity:
\begin{equation}
\label{e:Modulus}
\mu_g:\IN\to\IN \quad\text{minimal s.t.:}\quad 
|x-x'|\leq2^{-\mu(n)}\;\Rightarrow\; |g(x)-g(x')|\leq2^{-n} \enspace .
\end{equation}
It roughly means that $\approx\mu_g(n)$ bits of the input approximation are sufficient and necessary
in order to guarantee $\approx n$ correct bits of the output approximation.
\begin{enumerate}
\item[a)]
If $g$ can be computed (i.e. its output value approximated up to error $2^{-n}$)
in time $t(n)$, then its modulus of continuity satisfies $\mu_g(n)\leq t(n+2)$ \cite[Theorem~2.19]{Ko91}.
\item[b)]
Conversely, if $g$ has modulus of continuity $\mu_g$, 
then there exists some (oracle and an) oracle Turing machine 
computing $g$ in time $\calO\Big(\mu_g\big(\calO(n)\big)\Big)$
\cite[Corollary~3.6b]{LimZiegler25}.
\item[c)]
\todo[inline]{Should it be $g\in\calC^\alpha([0;1])$ and $\mu(n)\leq\calO(n^\alpha)$?}\todo[inline]{YES for the `unary' conception of modulus, NO for our `binary' modulus: see the new Example~\ref{x:Modulus2}}
A function $g\in\calC([0;1])$ is H\"{o}lder-continuous ~iff~
its modulus of continuity is linearly bounded: $\mu_g(n)\leq\calO(n)$.
It is Lipschitz-continuous ~iff~
its modulus of continuity satisfies $\mu_g(n)\leq n+\calO(1)$.
\item[d)]
The function $h:[0;1]\to [0;1], \;h(t)=1/\ln(e/t)$ 
has exponential modulus of continuity.
\item[e)]
Iterate $h\circ h:[0;1]\to [0;1]$ has \emph{doubly} exponential modulus of continuity, and so on.
\item[f)]
If $f$ has modulus $\mu$ and $g$ has modulus $\nu$, then $g\circ f$ has modulus $\leq\mu\circ\nu$.
\item[g)]
If $g_m$ has modulus $\mu_m$ and if it holds $\|f-g_m\|_\infty:=\sup_x |f(x)-g_m(x)|\leq 2^{-m}$,
then $f$ has modulus $\mu_f(n)\leq \mu_{n+2}(n+1)$.
\end{enumerate}
\end{example}
See Example~\ref{x:Modulus2} for other conceptions of ``modulus of continuity''.

\subsubsection[Parameterized Complexity on $\sigma$-Compact Domains]{Parameterized Complexity on $\pmb\sigma$-Compact Metric Domains:}
\label{sss:SigmaCompact} 
De\-fining the complexity of a function $g:X\to\IC$
becomes more subtle when its domain $X$ is merely $\sigma$-compact.  
Then one can express $X$ as a countable monotone cover $X=\bigcup\nolimits_{m\in\IN} X_m$, 
each piece $X_m\subseteq X_{m+1}$ being compact.
This naturally leads to a \emph{parameterized} conception of time complexity $t(n,m)$.  
Like in the discrete case, one commonly hopes for time \emph{fully} polynomial in $(n+m)$;
or at least \emph{fixed-parameter tractability}, meaning time bound $\phi(m)\cdot n^{\calO(1)}$
for some arbitrary non-decreasing $\phi:\IN\nearrow\IN$ \cite{FlumGrohe}.
Again, changing the cover can yield dramatically different complexity behaviors:

\begin{example}
\label{x:Exp}
A classic example is $\exp : \IR \to \IR$.
Cover $X_m=[-m;+m]$ allows computing $\exp$ in time polynomial in $(n+m)$
by evaluating the first $K=\calO(n+m)$ terms of the Taylor expansion $\sum_{k=0}^{K} t^k/k!$;
whereas cover $X'_m=[-2^m;+2^m]$ incurs time complexity polynomial only in $(n + 2^m)$,
again for information-theoretic reasons 
since on argument $x:=2^m$ already the integer part ($n=0$) 
of the output occupies $\approx2^m$ bits.
\end{example}

\subsection{Parameterizing the Space of Continuous Functions}
\label{ss:Continuous}
When one studies functionals (or operators) in Analysis---receiving as arguments real or complex functions 
instead of numbers or finite-dimensional vectors---further delicacies arise \cite{KC12}.
For instance the space $\calC(X,R)$ of all continuous functions $f:X\to\IC$ with $|f(x)|\leq R$,
as candidate domain of an operator, is \emph{not} generally $\sigma$-compact.
This subsection reviews several common \emph{second-order} (Subsubsection~\ref{sss:SecondOrder}) parameterizations,
arising for example from best approximation polynomial degree
(Subsubsection~\ref{sss:Uniform}), together with 
their justifications (Subsubsection~\ref{sss:Entropy})
as well as refutations of other parameterizations 
(Subsubsection~\ref{sss:Skolem}).
For a start, the Arzel\`{a}-Ascoli Theorem 
\cite[Theorem~4.25]{Brez11FASS}
suggests a natural parameterization and cover of $\calC(X)$ by compact subsets:

\begin{example}[Quantitative Arzel\`{a}-Ascoli]
\label{x:Arzela}
Fix a compact metric space $(X,d)$.
Then any closed and bounded and equicontinuous subset of $\calC(X)$ is compact.
\begin{enumerate}
\item[a)]
In particular,
for any fixed non-decreasing $\mu:\IN\nearrow\IN$ and any $R\geq0$, 
the following subset is compact:
\[ \calC_{\mu}(X,R) \;:=\; \big\{ g:X\to\IC \text{ has modulus of continuity } \leq\mu,  |g(x)|\leq R \big\} \enspace . \]
\item[b)]
For $R\leq T$ and $\mu\leq\nu$ it holds $\calC_{\mu}(X,R)\subseteq\calC_{\nu}(X,T)$.
Moreover $R<T$ implies $\calC_{\mu}(X,R)\subsetneq\calC_{\mu}(X,T)$.
If $X$ is furthermore infinite, then $\mu\neq\nu$ implies $\calC_{\mu}(X,R)\neq\calC_{\nu}(X,R)$.
\item[c)]
If $\calC'\subseteq\calC(X)$ is compact,
then there exists some minimal $\mu:\IN\nearrow\IN$ and $R\geq0$
such that $\calC'\subseteq\calC_{\mu}(X,R)$.
\end{enumerate}
\end{example}
This suggests the monotone compact cover 
\begin{equation}
\label{e:Arzela}
\calC(X) \quad=\quad \bigcup\limits_{\substack{r\in\IN \\ \mu\in\IN\nearrow\IN}} \calC_\mu(X,2^r) \enspace . 
\end{equation}
Note that it suffices to consider only countably many radii,
but $\mu$ cannot be restricted to any countable subset of $(\IN\!\nearrow\!\IN)$ 
according to Example~\ref{x:Arzela}b).

\subsubsection{Degree-Rate of Weierstra\ss{} Approximation:}
\label{sss:Uniform}
Univariate polynomials with complex coefficients are dense in $\calC([0;1])$ according to Weierstra\ss.
Let us call $\delta=\delta_f:\IN\nearrow\IN$ the \emph{degree-rate of Weierstra\ss{} approximation} 
to $f\in\calC([0;1])$
if there exist polynomials $p_n\in\IC[X]$ of $\deg(p_n)\leq2^{\delta(n)}$,
but not of $\deg(p_n)\leq2^{\delta(n)-1}$, such that it holds
$\|f-p_n\|_\infty\leq2^{-n}$.
Classical Approximation Theory asserts that $\delta_f$ and the modulus of continuity $\mu_f$ are linearly related
in argument and value, as summarized in Fact~\ref{f:JacksonMarkov}d).
Items~(a)+(b) first recall the Jackson III Theorem and Markov Brothers' Inequality \cite[\S4.6+\S3.7]{Cheney}.

\begin{fact}
\label{f:JacksonMarkov}
\begin{enumerate}
\item[a)]
Suppose $g:[0;2\pi]\to\IR$ has `continuous' modulus of continuity $\omega$ in the sense of Example~\ref{x:Modulus2}d).
Then $g$ can be uniformly approximated up to error $\omega\big(\tfrac{\pi}{D+1}\big)$
by some polynomial $P\in\IR[X]$ of degree $\leq D$.
\item[b)]
For any polynomial $P\in\IR[X]$ of degree $\leq D$, it holds
\[ \max\big\{ |P'(x)| : -1\leq x\leq+1 \big\} \;\leq\; 
D^2 \cdot\max\big\{ |P(x)| : -1\leq x\leq+1 \big\} \enspace . \]
If $|P(x)|\leq2$ holds for all $x\in[-1;+1]$, 
then $P$ has modulus $\mu(n)\leq n+1+2\Log(D)$. 
\item[c)]
If $f:[-1;1]\to[-1;+1]$ can be uniformly approximated up to error $2^{-m}$ 
by $P_m\in\IR[X]$ of degree $D_m$,
then $f$ has modulus $\mu_f(n)\leq n+2+2\Log(D_{n+2})$.
\item[d)]
Any $f\in\calC([0;1],1)$ satisfies
\[ \delta_f(n) \;\;\leq\;\; \mu_f\big(n+\calO(1)\big)+\calO(1) 
\;\;\leq\;\;
n+\calO(1)+2\delta_f\big(n+\calO(1)\big) \enspace . \]
\end{enumerate}
\end{fact}
Fact~\ref{f:JacksonMarkov}d) is a prototype for our investigations
of compact parameterizations of $\ELL{p}$ in Theorem~\ref{t:Equivalent}.
Fact~\ref{f:JacksonMarkov}d) combines (a) and (c);
and c) in turn follows from (b) with Example~\ref{x:Modulus}g).

\subsubsection{Metric Entropy:}
\label{sss:Entropy}

We follow \cite{KSZ16a,LZ20} in 
adopting Kolmogorov's metric entropy \cite{Kolmogorov,Wei03,Koh08a} 
to the `binary' paradigm, elaborated in Subsubsection~\ref{sss:Skolem} below,
capturing total boundedness as follows:

\begin{definition}
\label{d:Entropy}
The (binary) \emph{entropy} of compact metric space $(X,d)$ 
is the non-decreasing mapping $\Entropy=\Entropy_X:\IN\to\IN$ 
such that $X$ can be covered by $2^{\Entropy(n)}$, but not by $2^{\Entropy(n)-1}$,
closed balls of radius $2^{-n}$.
\end{definition}
See also \cite[Lemma~2.3+Example~2.6]{LimZiegler25}. 
We record in particular \cite[Theorem~5.11]{Ste17}: 

\begin{fact}
\label{f:Entropy}
The space $\calC_\mu([0;1],1)$ of mappings $g:[0;1]\to\calD$
with modulus of continuity $\mu_g\leq\mu$
has entropy $2^{\mu(n\pm\calO(1))\pm\calO(1)}$.
\end{fact}
\noindent
Note that entropy is a quantitative property of 
sets of, rather than of individual, functions.
It therefore does not by itself induce a cover of $\calC([0;1],1)$.

\subsubsection{Second-Order Complexity Viewpoint:}
\label{sss:SecondOrder}
Equation~\eqref{e:Arzela} naturally leads to a conception of time complexity $t(n,r,\mu)$
involving, in addition to first-order integer parameters $n$ and $r$, 
the \emph{second-order} parameter $\mu:\IN\nearrow\IN$.
Again one commonly hopes for time bounded by some (now also second-order) polynomial in $n,r,\mu$
\cite{DBLP:journals/ndjfl/Townsend90,KC96,KC12,lim2023degreessecondhigherorderpolynomials}.

Like a classical (=first-order) polynomial bound allots more time to an algorithm 
on longer discrete inputs or for computing approximations with more reliable output bits,
second-order polynomial bounds automatically allot more time to an algorithm computing 
a functional or operator on function arguments with larger modulus of continuity---justified
by Example~\ref{x:Modulus}a).
Specifically, the application functional 
\begin{equation}
\label{e:Application}
\calC(X,1)\times X \;\ni\; (g,x) \;\mapsto\;  g(x)\;\in\;\calD\;\subseteq\;\IC 
\end{equation}
is second-order polynomial-time computable by a suitable oracle Turing machine
\cite[Lemma~4.9]{KC12},
where $\calD\subseteq\IC$ denotes the complex unit disc.

\subsubsection{Binary versus Unary Skolem Moduli:}
\label{sss:Skolem}

Mathematical Logic suggests \emph{Skolemization} as
a generic way of capturing qualitative $\Pi_2=\forall\exists$ statements 
quantitatively in terms of non-decreasing number-theoretic mappings $\phi:\IN\nearrow\IN$ aka \emph{moduli}; 
cmp. \cite[Ex~4.8.2+Def~17.106+p.285+\S15.4+p.379+Def~17.116]{Koh08a}.
Skolemizing \emph{uniform continuity} ``$\forall\epsilon>0\exists\delta>0$'' 
recovers the \emph{modulus of continuity} $\mu$ from Example~\ref{x:Modulus};
and Skolemizing total boundedness of a metric space 
yields the \emph{entropy} $\Entropy$ in Subsubsection~\ref{sss:Entropy}.

Skolemizing a $\forall\exists$ statement as modulus $\phi:\IN\nearrow\IN$ can be done in many ways.
Two major approaches may be called \emph{unary} and \emph{binary}, respectively.
Let us illustrate them with the following variant of Example~\ref{x:Modulus}:

\begin{example}
\label{x:Modulus2}
Fix a uniformly continuous function $f:X\to Y$ between metric spaces $(X,d)$ and $(Y,e)$.
Its \emph{unary} modulus of continuity 
is $M=M_f:\IN_+\nearrow\IN_+$ minimal such that
$d(x,x')\leq1/M(N)$ implies $e\big(f(x),f(x')\big)\leq1/N$.
\begin{enumerate}
\item[a)]
Suppose $X$ is compact or complete convex or $Y$ is bounded.
Then $f$ is Lipschitz-continuous ~iff~ its unary modulus satisfies $M_f(N)\leq\calO(N)$.
And $f$ is H\"{o}lder--continuous ~iff~ its unary modulus satisfies $M_f(N)\leq\poly(N)$.
\item[b)]
If $M$ is the unary modulus of $f$, then its binary modulus exists and satisfies $\mu_f(n)\leq\Log M(2^n)$,
where $\Log(K):=\lceil\log_2(K)\rceil$.
\item[c)]
If $\mu$ is the binary modulus of $f$, then its unary modulus exists and satisfies $M_f(N)\leq2^{\mu(\Log N)}$.
\item[d)]
A further, `continuous', conception of modulus of continuity from Calculus:
\[ \omega_g:[0;1]\to[0;\infty), \quad \omega_g(\delta)\;=\;\max\big\{e\big(g(x),g(x')\big):e(x,x')\leq\delta\big\} \enspace . \]
From that one can recover both
unary $M_g(N)=\min\big\{M:\omega_g(1/M)\leq1/N\big\}$ and 
binary $\mu_g(n)=\min\big\{m:\omega_g(2^{-m})\leq2^{-n}\big\}$.
\end{enumerate}
\end{example}
\begin{remark}
\label{r:Binary}
\begin{enumerate}
\item[a)]
The literature also knows `mixed' unary/binary conceptions of moduli.
For example \cite[Definition~18.52]{Koh08a} considers as modulus of \emph{total boundedness}
$\Gamma:\IN\nearrow\IN$ such that the space
can be covered by $\Gamma(n)$, but not by less,
closed balls of radius $2^{-n}$; 
\cite[\S6]{Wei03} calls this \emph{width}.
The (`purely') binary Definition~\ref{d:Entropy} is then recovered as $\Entropy=\Log\Gamma$.
\item[b)]
In the discrete case, binary (as opposed to unary) encoding is common;
recall Subsubsection~\ref{sss:Discrete}. 
Accordingly, we favor in this work the binary 
conception of the modulus of continuity from Example~\ref{x:Modulus}a+b);
and similarly and in turn \cite[Lemma~2.3c]{LimZiegler25} the binary entropy.
\item[c)]
Numerics does consider both unary $\varepsilon=1/N$ and binary $\varepsilon=2^{-n}$ error bounds.
The latter has arguably several structural advantages:
\item[c\,i)]
Famous algorithms and implementations approximating $\pi$ up to guaranteed error $\varepsilon\approx 10^{-10^{12}}$ {\rm\cite{pi}}
are appreciated for running in time `polynomial in $\log(1/\varepsilon)\approx 3.3\cdot10^{12}$', 
rather than merely in time `polynomial in $1/\varepsilon=10^{10^{12}}$'.
\item[c\,ii)]
Common numerical problems are easily seen to support approximations
up to error $1/n$ in time polynomial in $n$, 
whereas doing so up to error $2^{-n}$
has been shown equivalent to famous open conjectures regarding 
discrete complexity classes 
like $\classFP$ versus $\sharpP$ or $\classFPSPACE$;
cmp. {\rm\cite{Ko91,Kaw10,Ste17,Boche,KPSZ23}}.
\item[c\,iii)]
The first-order logic of the integers is well-known 
incomplete/undecidable since G\"{o}del; 
that of the reals on the other hand is logically
complete/decidable according to Tarski. 
However expanding the reals with a `unary' integer embedding like
$\IN_+\ni n\mapsto 1/n\in\IR$ recovers incompleteness;
but expanding the reals instead with the `binary' integer embedding
$\IZ\ni n\mapsto 2^{-n}\in\IR$ does preserve logical completeness
\cite[\S6.1]{Sewon24}.
\item[d)]
In agreement with its name, the exponential function arguably 
\emph{should} have exponential complexity. 
According to Example~\ref{x:Exp}, this suggests 
the parameterized `binary' cover $\IR=\bigcup\limits_m [-2^m;+2^m]$
rather than `unary' $\IR=\bigcup\limits_m [-m;+m]$.
\item[e)]
Similarly the parameterized cover of $\calC(X)$ 
using binary moduli of continuity according to Equation~\eqref{e:Arzela}
is preferable over, say, `unary' parameterization
\[
\calC(X) \quad=\quad \bigcup\limits_{\substack{R\in\IN, \\ M\in\IN_+\nearrow\IN_+}} 
\big\{ g:X\to\IC \text{ has \emph{unary} modulus } \leq M,  |g(x)|\leq R \big\}
\enspace . \]
\end{enumerate}
\end{remark}
In conclusion, the present work focuses on `binary' conceptions:
also for moduli/rates and the induced parameterizations of $\ELL{p}$ in 
Subsections~\ref{sss:Uniform}, \ref{ss:LpModulus}, \ref{ss:StepModulus}.
That said, binary and unary estimates can easily be translated back and forth
similarly to Example~\ref{x:Modulus2}b+c).

\medskip
\begin{enumerate}
\item[i)]
We try to reserve lowercase letters like $n,m,r$ 
for unbounded `binary' natural number parameters,
occurring in the exponent to base $2$;
and uppercase letters like $N=2^{n}$ or $R=2^r$
for their `unary' counterparts.
\item[ii)]
For quantities tending to zero, classically typically denoted $\varepsilon$ or $\delta$, 
we use lowercase letters as `binary' natural number parameters
occurring as exponents to base $\tfrac{1}{2}$,
like $\varepsilon=2^{-n}$ or $\delta=2^{-\mu(n)}$.
\item[iii)]
Similarly we use lowercase Greek letters like $\mu,\Entropy$
to denote moduli with both `binary' arguments and values;
uppercase Greek letters denote moduli with both `unary' arguments and values.
\item[iv)]
Like computational complexity analyses, 
also estimates of and among moduli are often more robustly/concisely
stated (for instance covering both real and the complex valued functions
over both $\IS$ and over $\IT$) 
by neglecting constants using Landau's $\calO()$-notation.
\item[v)]
Here we shall thus often ignore constant offsets in binary arguments and/or values 
of moduli: which amounts to ignoring constant factors
in unary arguments/values:
such as in Fact~\ref{f:Entropy} above, or in 
Equation~\ref{e:Embeddings2} or in Fact~\ref{f:JacksonMarkov}d) below.
\item[vi)]
Beyond (v), we may occasionally ignore as well 
constant factors---instead of offsets---in binary parameters and values;
which amounts to ignoring polynomial dependencies in their unary counterparts.
\end{enumerate}

\section{Spaces of Integrable Functions}
\label{s:Integrable}
Similarly to the case of continuous functions (Subsection~\ref{ss:Continuous}),
this work establishes and justifies a second-order parameterization of spaces
of integrable functions. More precisely we consider the space 
$\ELL{p}(\IT)$ of equivalence classes of Lebesgue measurable and $p$-integrable functions on the unit interval with wraparound $\IT=[0;1)\bmod 1$ aka ``real unit circle'' for any $1<p<\infty$. Recall the $p$-norm $\|f\|_p=\sqrt[p]{\int |f(x)|^p\,dx}$.

Like $\calC(\IT)$, $\ELL{p}(\IT)$ is not $\sigma$-compact---and 
$\ELL{\infty}(\IT)$ not even separable. 
For $1<p<q<\infty$ record embeddings
\begin{equation}
\label{e:Embeddings}
\calC(\IT) \;\subsetneq\; \ELL{\infty}(\IT) \;\subsetneq\; 
\ELL{q}(\IT) \;\subsetneq\; 
\ELL{p}(\IT) \;\subsetneq\; 
\ELL{1}(\IT) , \quad 
\|g\|_p\leq \|g\|_q \enspace .
\end{equation}
Also note that, unlike for $g\in\calC(\IT)$, 
point evaluation $\IS\ni t\mapsto g(t)\in\IC$ 
is in general not well-defined for $g\in\ELL{p}(\IT)$.
Hence criteria for/against some choice of second-order parameter 
based on Example~\ref{x:Modulus}a+b),
similarly to the modulus of continuity for $\calC(X)$
in Subsection~\ref{ss:Continuous},
cannot carry over to $\ELL{p}(\IT)$.

Instead we consider several second-order parameterizations of $\ELL{p}(\IT)$
arising naturally from common mathematically equivalent approaches:
including $\ELL{p}$-modulus \cite{Ste17} in Subsection~\ref{ss:LpModulus},
rate of approximation by step functions in Subsection~\ref{ss:StepModulus},
and rate of approximation by Fourier series in Subsections~\ref{ss:FourierRate}.
Our main result in Subsection~\ref{ss:Equivalent} establishes them mutually linearly related,
and thus arguably all (almost equally) `reasonable' from a polynomial complexity perspective.
The technical proofs are postponed to Subsection~\ref{ss:Proof}.

\subsection[{Lp}-Modulus]{$\Lrm{p}$-Modulus}
\label{ss:LpModulus}

This subsection recalls from \cite{Ste17}
the conception of the (binary) $\Lrm{p}$-modulus
as Skolemization of the Fr\'{e}chet--Kolmogorov Theorem,
and some first justification of this modulus
as suitable second-order parameterization of $\ELL{p}(\IT)$.

\begin{definition}[$\Lrm{p}$-modulus]
\label{d:LpModulus}
Let $\IT=[0;1)\bmod 1$ denote the real unit interval with 
`wraparound', and fix $p \ge 1$. 
We say that $\mu=\mu^{(p)}_f:\IN \nearrow \IN$ is the (binary) $\Lrm{p}$-modulus
of $f \in \Lrm{p}(\IT)$ iff $\mu$ is pointwise minimal s.t.
\[
  \big\|f - \tau_\delta f\big\|_{p} \; \le \; 2^{-n}
  \quad 
  \text{for all $|\delta|\le 2^{-\mu(n)}$.}
\]
Here, $\big(\tau_\delta f\big)(t) := f(t+\delta \bmod 1)$ denotes the shift of $f$ by $\delta$ with wraparound.
Moreover, we denote by $\Lrm{p}_{\mu}(\IT,R)\subseteq \Lrm{p}(\IT)$
the set of all $f$ bounded by $\|f\|_p \le R$ and with $\Lrm{p}$-modulus $\mu^{(p)}_f\leq\mu$.
\end{definition}
Analogously to the Arzel\`{a}-Ascoli Theorem (Example~\ref{x:Arzela}) 
for spaces $\calC(X)$ of continuous functions,
the Fr\'{e}chet--Kolmogorov Theorem describes compact subsets 
of the spaces $\ELL{p}(\IT)$ of $p$-integrable functions:

\begin{fact}[Quantitative Fr\'{e}chet--Kolmogorov]
\label{f:Frechet}
Fix $p\geq1$.
\\ Every $f\in\ELL{p}(\IT)$ has an $\ELL{p}$-modulus $\mu_f$ 
{\rm\cite[Theorem~4.26]{Brez11FASS}}.
\begin{enumerate}
\item[a)]
For non-decreasing $\mu:\IN\nearrow\IN$ and $R\geq0$, 
$\Lrm{p}_{\mu}(\IT,R)$ is compact.
\item[b)]
For $R\leq T$ and $\mu\leq\nu$ it holds $\ELL{p}_{\mu}(\IT,R)\subseteq\ELL{p}_{\nu}(\IT,T)$.
Moreover $R<T$ implies $\ELL{p}_{\mu}(\IT,R)\subsetneq\ELL{p}_{\mu}(\IT,T)$.
Furthermore $\mu\neq\nu$ implies $\ELL{p}_{\mu}(\IT,R)\neq\ELL{p}_{\nu}(\IT,R)$.
\item[c)]
If $\ELL{'}\subseteq\ELL{p}(\IT)$ is compact,
then there exists some minimal $\mu:\IN\nearrow\IN$ and $R\geq0$
such that $\ELL{'}\subseteq\ELL{p}_{\mu}(\IT,R)$.
\item[d)]
For $1\leq p\leq q\leq \infty$ and
$f\in\ELL{q}(\IT)$, it holds
$\mu_f^{(p)}\leq \mu_f^{(q)}$
according to Embedding~\eqref{e:Embeddings}.
\item[e)]
Similarly to Example~\ref{x:Modulus}g),
if $g_m\in\ELL{p}(\IT)$ have $\ELL{p}$-modulus $\mu_m$
and if $\|f-g_m\|_p\leq2^{-m}$, then $f\in\ELL{p}(\IT)$ and has 
$\ELL{p}$-modulus $\mu_f(n)\leq \mu_{n+2}(n+1)$.
\end{enumerate}
\end{fact}
This suggests the monotone compact cover 
\begin{equation}
\label{e:Frechet}
\ELL{p}(\IT) \quad=\quad \bigcup\limits_{\substack{r\in\IN,\\ \mu\in\IN\nearrow\IN}} \ELL{p}_\mu(\IT,2^r) \enspace , 
\end{equation}
and similarly for $\ELL{p}(\IS)$:

\begin{remark}
\label{r:CircleVsInterval}
We straightforwardly rephrase Definition~\ref{d:LpModulus} and Fact~\ref{f:Frechet}
for the $\Lrm{p}$-modulus $\mu_g$ of a $p$-integrable function $g$
on the complex unit circle $\IS$ instead of the real $\IT$.
Since $\IS$ has length $2\pi\in[2^2;2^3]$ instead of length $1$ for $\IT$,
$g\big(\exp(2\pi i t)\big)\equiv f(t)$ 
has $\mu_f+3\leq \mu_g\leq \mu_f+2$.
Over $\IS$ instead of $\IT$, 
the last estimate in Equation~\eqref{e:Embeddings} 
and Fact~\ref{f:Frechet}d) now become:\todo{I think the same estimate should still be true for the unit circle S, no?}
\begin{equation}
\label{e:Embeddings2}
\|g\|_p \;\leq\; \calO\big(\|g\|_q\big) \quad , \qquad
\mu^{(p)}_g\big(n\big)\;\leq\;\mu^{(q)}_g\big(n+\calO(1)\big)
\end{equation}
with the constants in $\calO()$ depending on 
$p\leq q$ but not on $g\in\ELL{q}(\IS)$.
\end{remark}
\begin{remark}
\label{r:Florian}
\begin{enumerate}
\item[i)]
\cite[\S3.1]{Ste17} had defined $\ELL{p}$-modulus for integrable functions
$f$ on the real unit interval $[0;1]$ with\emph{out} wraparound $\IT$.
There, a shifted $\tau_\delta f$ is considered re-extended to entire $[0;1]$
with value $\equiv0$. 
\item[ii)]
\cite[\S5.3]{Ste17} had justified parameterizing $\ELL{p}$ 
by the $\ELL{p}$-modulus according to Equation~\ref{e:Frechet}
because the Fr\'{e}chet--Kolmogorov Fact~\ref{f:Frechet}
is generally considered the $\ELL{p}$-analogon to---and indeed commonly 
proven based on---the 
Arzel\`{a}-Ascoli Example~\ref{x:Arzela} for $\calC$.
\item[iii)]
Moreover, \cite[Theorem~5.24]{Ste17} establishes
that $\ELL{p}_\mu(\IT,1)$ has entropy \linebreak[4]
$2^{\calO(\mu(n\pm\calO(1)))}$: 
similarly to Fact~\ref{f:Entropy} for $\calC_\mu([0;1],1)$.
\end{enumerate}
\end{remark}

\subsection{Rate of Approximation by Step Functions}
\label{ss:StepModulus}

Recall Subsection~\ref{sss:Uniform} parameterizing $\calC([0;1])$
via degree-rates of uniform approximation by polynomials,
the latter known dense in $\calC([0;1])$ according to Weierstra\ss.
Similarly, \emph{step functions} are a
classically dense subset of $\ELL{p}(\IT)$.

\begin{definition}
\label{d:StepRate}
For $K\in\IN$ let $\calS_{K}(\IT)$ denote the vector space of complex functions
which are constant on each of the $K$ half-open subintervals 
$\big[k/K;(k+1)/K\big)\subseteq\IT$, $0\leq k<K$.

The (binary) \emph{rate of piecewise approximation} (``Step rate'')
$\sigma^{(p)}_f$ of $f \in \Lrm{p}(\IT)$
is the pointwise least $\sigma:\IN\nearrow\IN$ such that
there exists some $f_n\in\calS_{2^{\sigma(n)}}(\IT)$ 
with $\|f-f_n\|_p\leq2^{-n}$.
\end{definition}
Similarly to Remark~\ref{r:CircleVsInterval},
we consider the step rate $\sigma^{(p)}_g:=\sigma^{(p)}_f\pm\calO(1)$ 
of piecewise approximation also of $g \in \Lrm{p}(\IS)$ 
with $g\big(\exp(2\pi i t)\big):\equiv f(t)$.

\subsection{Rate of Fourier Series Convergence}
\label{ss:FourierRate}

Recall the Fourier coefficients and Fourier sums
associated with any integrable periodic function $f\in\ELLone(\IT)$:
\begin{equation}
\label{e:Fourier}
\hat f_k =\; \int\limits_0^1 f(t)\cdot\exp(-2\pi i k t)\,dt \quad (k\in\IZ) 
\; , \quad 
\calF_K f(t) =\; \sum\limits_{|k|\leq K} \hat f_k\cdot \exp(2\pi i k t)
\end{equation}
It is known that $\calF_K f$ converges to $f\in\ELL{p}(\IT)$ in $p$-norm when $1<p<\infty$,
but not necessarily for $p=1$ nor $p=\infty$ \cite[\S1.4]{Duoandikoetxea};
see the below proof of Theorem~\ref{t:Equivalent}a)
for a quantitative analysis in terms of the following parameter:

\begin{definition}
\label{d:FourierRate}
The (binary) \emph{rate of Fourier series convergence} (``Fourier rate'')
$\varphi_f^{(p)}$
of $f\in\ELL{p}(\IT)$
is the pointwise least $\varphi:\IN\nearrow\IN$ 
with $\displaystyle\big\|f-\calF_{2^{\varphi(n)}} f\big\|_p\leq2^{-n}$.
We denote by $\varphi=\varphi_f$ also the Fourier rate 
of $g\in\ELL{p}(\IS)$, $g\big(\exp(2\pi i t)\big)\equiv f(t)$.
\end{definition}

\subsection{Main Result: Equivalence of the above Parameters}
\label{ss:Equivalent}

Theorem~\ref{t:Equivalent} establishes that, for $1<p<\infty$, 
the above three second-order parameterizations
of $\ELL{p}$ are `linearly' equivalent.
It suffices to treat functions with complex unit disc as co-domain,
since scaling extends to functions with values bounded by $2^r$:
\begin{equation}
\label{e:Scaling}
\mu_{f\cdot 2^r}(n)=\mu_f(n+r), \qquad
\sigma_{f\cdot 2^r}(n)=\sigma_f(n+r), \qquad
\varphi_{f\cdot 2^r}(n)=\varphi_f(n+r) \enspace . 
\end{equation}

\begin{theorem} 
\label{t:Equivalent}
Fix $1<p<\infty$ and, for $f\in\Lrm{p}(\IT,1)$,
recall the $\ELL{p}$-modulus $\mu=\mu^{(p)}_f$ from Definition~\ref{d:LpModulus} and
the Step rate $\sigma=\sigma^{(p)}_f$ from Definition~\ref{d:StepRate}
and the Fourier rate $\varphi=\varphi^{(p)}_f$ from Definition~\ref{d:FourierRate}.
Then it holds:
\begin{enumerate}
\item[a)] $\displaystyle
\varphi(n)\;\leq\; \mu\big(n+\calO(1)\big)\:+\:n\:+\:\calO(1)$.
\item[b)] $\displaystyle
\mu(n) \;\leq\; 2\varphi\big(n+\calO(1)\big)\:+\:n\:+\:\calO(1)$.
\item[c)] $\displaystyle
\sigma(n)\;\leq\; \mu\big(n+\calO(1)\big)$.
\item[d)] $\displaystyle
\mu(n)\;\leq\; \sigma\big(n+\calO(1)\big)\:+\:pn\:+\:\calO(1)$.
\end{enumerate}
Here constants hidden in $\calO(1)$ may depend on $p$ but not on $f$.
\end{theorem}
\noindent
The proof of Theorem~\ref{t:Equivalent} is deferred to Subsection~\ref{ss:Proof}.

\begin{corollary}
\label{c:Equivalent}
For fixed $1<p<\infty$ and for all $f\in\Lrm{p}(\IT,1)$, it holds
\[
\calO\Big(\varphi^{(p)}_f\big(n+\calO(1)\big)\;+\;n\Big)
\;=\;
\calO\Big(\mu^{(p)}_f\big(n+\calO(1)\big)\;+\;n\Big)
\;=\;
\calO\Big(\sigma^{(p)}_f\big(n+\calO(1)\big)\;+\;n\Big)
\]
and same for $g\in\Lrm{p}(\IS,1)$.
Again, 
constants hidden in $\calO()$ may depend on $p$ but not on $f$ nor $g$.
\end{corollary}
Recall Subsubsection~\ref{sss:Skolem} that this applies to 
`binary' moduli/rates. For their `unary' counterparts,
Theorem~\ref{t:Equivalent} yields polynomial equivalence:
similarly to Example~\ref{x:Modulus2}b+c).

\subsection{Proof of Theorem~\ref{t:Equivalent}}
\label{ss:Proof}

Record some folklore estimates among $p$-norms 
of vectors, sequences, and functions for $p\geq q\geq1$,
where $p':=\tfrac{p}{p-1}$:
\begin{gather}
\label{e:Hoelder0}
(|x|+|y|)^p \;\leq\; 2^{p-1} (|x|^p+|y|^p) 
\\
\label{e:Hoelder1}
\sum\nolimits_n |x_n\cdot y_n|
\;\leq\; 
\Big( \sum\nolimits_n |x_n|^p \Big)^{1/p} \:\cdot\:
\Big( \sum\nolimits_n |y_n|^{p'}  \Big)^{1/p'} 
\\
\label{e:Embeddings3} 
\Big(\sum\limits_{k=0}^{K-1} |x_k|^p\Big)^{1/p}
\;\;\leq\;\;
\Big(\sum\limits_{k=0}^{K-1} |x_k|^q\Big)^{1/q}
\;\;\leq\;\;
K^{\tfrac{1}{q}-\tfrac{1}{p}}\cdot
\Big(\sum\limits_{k=0}^{K-1} |x_k|^p\Big)^{1/p}
\\
\label{e:Hoelder2} 
\int\nolimits_X |f(x)\cdot g(x)|\,dx \;\leq\; 
\Big( \int\nolimits_X |f(x)|^p \,dx \Big)^{1/p} \:\cdot\:
\Big( \int\nolimits_X |g(x)|^{p'} \,dx \Big)^{1/p'} 
\\
\label{e:Minkowski}
\Big( \int\nolimits_Y \Big| \int\nolimits_X F(x,y)\, dx\Big| ^{p} \, dy\Big)^{1/p}
\;\leq\; 
\int\nolimits_X \Big( \int\nolimits_Y |F(x,y)|^p \, dy \Big) ^{1/p} \, dx
\end{gather}
Inequality~\eqref{e:Hoelder0} follows from convexity,
\eqref{e:Minkowski} is Minkowski's Integral Inequality,
\eqref{e:Hoelder1} and \eqref{e:Hoelder2} are H\"{o}lder's---which 
also imply \eqref{e:Embeddings3} and \eqref{e:Embeddings2}, respectively.

\begin{enumerate}
\item[a)]
Recall the Ces\`{a}ro average 
$\sum_{k=0}^{K} \calF_k f/(K+1)=f\ast F_K$ 
of $f$'s first $(K+1)$ Fourier sums
$\calF_K f= f\ast D_K$ 
as convolutions with the Dirichlet and Fej\'{e}r kernels
\begin{eqnarray*}
D_K(t) &:=&\sum\nolimits_{k=-K}^{+K} \exp(2\pi i k t \;=\; \frac{\sin\big(\pi (2K+1) t\big)}{\sin(\pi t)} \qquad , \\
F_{K}(t) &:=& \sum\nolimits_{k=0}^{K} D_k(t)/(K+1) \;=\; \left( \frac{\sin\big(\pi (2K+1) t\big)}{\sin(\pi t)} \right)^2 / (K+1) 
\;\geq0 \enspace . 
\end{eqnarray*}
Following the proof of \cite[Theorem~1.10]{Duoandikoetxea}, \todo{Ich glaube, dass dieser Beweis überflüssig ist, 
da wir ja eigentlich über die Bestapproximation (Theorem 6.1 und Corollary 6.2) eine bessere Abschätzung erhalten haben, oder? 
Diesen Teil der Arbeit hatte ich ja auch auskommentiert.}
since $\int F_K(t)\,dt=1$ and $F_K(-t)=F_K(t)$, 
\begin{eqnarray*}
\| f-f\ast F_K \|_p 
&=&
\bigg( \int\nolimits_0^1 
\Big| \int\nolimits_{-1/2}^{+1/2} | f(t-s)-f(t) |\cdot F_K(s)\,ds \Big|^p 
\,dt\bigg) ^{1/p} \\
&\overset{\eqref{e:Minkowski}}{\leq}&
\int\nolimits_{-1/2}^{+1/2} 
\Big( \int\nolimits_0^1 |f(t-s)-f(t)|^p \cdot F_K^p(s)\, dt
\Big)^{1/p} \, ds \\
&=& 
\int\nolimits_{-1/2}^{+1/2}
\| \tau_{-s} f -f \|_p \cdot F_K (s)\, ds \\
&\leq&
\int\nolimits_{|s|<\delta}
\underbrace{\| \tau_{-s} f -f \|_p} \cdot F_K (s)\, ds \;\;+\;\;
2\underbrace{\|f\|_p}\cdot \int_{|t|=\delta}^{1/2} F_K(s)\,ds  \\
&\leq& 2^{-(n+1)} \;+\; \calO\big(2^{\mu(n)}/(K+1)\big) 
\quad\leq 2^{-n}
\end{eqnarray*}
for $\delta:=2^{-\mu(n+1)}$ and $K\geq 2^{\mu(n+1)+n+\calO(1)}$
since 
$\int\nolimits_{\delta}^{1/2} F_K(s)\,ds \;=$
\[ =\; 
\int\limits_{\delta}^{1/2} \sin^{-2}(\pi s) \,ds / (K+1)
\;=\; 
-\frac{\cos(\pi s)}{\pi\sin(\pi s)\cdot(K+1)} \Big|_{s=\delta}^{s=1/2}
\;\leq\; \calO(1/\delta)/(K+1). \] 
It follows that
\[ 
\| \calF_K f- f \|_p
\;\leq\;
\| \calF_K (f- g_K) \|_p \;+\;
\| \calF_K g_K -g_K  \|_p \;+\;
\| f-g_K \|_p 
\;\leq\; (C_p+1)\cdot 2^{-n}
\]
for $g_K:=f\ast F_K$ (since $\calF_K g_K=g_K$)
and for the $(p,p)$ operator norm $C_p>0$ of the Dirichlet kernel multiplication operator,
which is known to be finite for $1<p<\infty$ (but not for $p=1$);
cmp. \cite[\S3.5]{Duoandikoetxea}.
\item[b)]
First record that
$|\exp(ir)-1|\leq |r|$ holds for all $r\in\IR$.
Next estimate the $\ELL{p}$-modulus of a finite Fourier series
$\displaystyle g_K=\sum\limits_{k=-K}^{+K} c_k \cdot\exp(2\pi ikt)$:

$\|g_K-\tau_\delta g_K\|_p^p$
\begin{eqnarray*}
&=&
\int\nolimits_0^1 
\Big| \sum\nolimits_{k=-K}^{+K} \exp(2\pi i k t) \cdot c_k \cdot  \big(1-\exp(2\pi i k \delta)\big) \Big|^p \, dt \\
&\overset{\eqref{e:Hoelder1}}{\leq}& 
\int\limits_0^1 
\Big( \underbrace{\sum\limits_{k=-K}^{+K} |\exp(2\pi i k t)|^{p'}}_{=2K+1} \Big)^{\tfrac{p}{p'}} 
\cdot 
\Big( \sum\limits_{k=-K}^{+K} |c_k|^p\cdot \underbrace{\big|1-\exp(2\pi i k \delta)\big|^p}_{\leq |2\pi k\delta|^p} \Big) \, dt \\
&\leq&
\big(2K+1\big)^{\tfrac{p}{p'}=p-1} \cdot \big| 2\pi K \delta\big|^p \cdot \|\vec c\|_p^p
\end{eqnarray*}
with the Fourier coefficient vector $\vec c\in\IC^{2K+1}$.
Then 
\[
|c_k| 
\;\overset{\eqref{e:Fourier}}{=}\;
\Big| \int\nolimits_0^1 g_K(t)\cdot\exp(-2\pi ikt)\,dt \Big| 
\;\leq\; 
\int\nolimits_0^1 |g_K(t)|\,dt 
\;=\; 
\|g_K\|_1 
\;\overset{\eqref{e:Embeddings2}}{\leq}\;
\|g_K\|_p 
\] 
and therefore 
\begin{equation}
\label{e:Estimate}
\|\vec c\|_p 
\;=\; \Big( \sum\nolimits_{k=-K}^{+K} |c_k|^p \Big)^{1/p} 
\;\leq\; \big(2K+1\big)^{1/p} \cdot \|g_K\|_p 
\enspace . \end{equation}
In conclusion,
\[
\|g_K-\tau_\delta g_K\|_p \;\leq\; 
(2K+1) \cdot | 2\pi K \delta| \cdot \|g_K\|_p
\quad\leq\;2^{-n}
\]
whenever $|\delta|\leq 2^{-\mu_K(n)}$
for $\mu_K(n):=n+\calO(1)+2\Log(K)$,
provided $\|g_K\|_p\leq2$.
With $K:=2^{\varphi(n)}$,
Fact~\ref{f:Frechet}e) yields the claim.
\\
PS: In case $p\geq2$ one could slightly improve Estimate~\eqref{e:Estimate} with \emph{Plancherel}:
\[ \|\vec c\|_p 
\;\overset{\eqref{e:Embeddings3}}{\leq}\;
\|\bar c\|_2 
\;=\; 
\|g_K\|_2 
\;\overset{\eqref{e:Embeddings2}}{\leq}\;
\|g_K\|_p 
\enspace . \]
\item[c)]
To $f\in\ELL{p}(\IT)$ and $K\in\IN$, define associated step functions 
$f_K\in\calS_K(\IT)$,
\[ f_K(t) \;:=\; K\cdot\int\limits_{k/K}^{(k+1)/K} f(s)\,ds 
\quad\text{for}\quad k/K\leq t<(k+1)/K,  \quad 0\leq k<K \enspace . \]
Then $\| f_K-f\|_p^p =$
\begin{eqnarray*}
&=& 
K^p \cdot \sum\nolimits_{k=0}^{K-1} \int\nolimits_{k/K}^{(k+1)/K} 
\Big| \int\nolimits_{k/K}^{(k+1)/K} 
1\cdot \big(f(s) - f(t)\big)\,ds \Big|^p \, dt \\
&\leq&
K^p \cdot
\sum\nolimits_{k=0}^{K-1} \int\nolimits_{k/K}^{(k+1)/K} 
\Big( \int\nolimits_{k/K}^{(k+1)/K} 1\cdot |\big(f(s) - f(t)\big|\,ds \Big)^p \, dt \\
&\overset{\eqref{e:Hoelder2}}{\leq}& 
K^p \cdot 
\sum\limits_{k=0}^{K-1} \int\limits_{k/K}^{(k+1)/K} 
\bigg( \Big( \int\limits_{k/K}^{(k+1)/K} 1^q\,ds \Big)^{1/q}
\cdot 
\Big(
\int\limits_{k/K}^{(k+1)/K} \big| f(s) - f(t) \big|^p\,ds\Big)^{1/p}  \bigg)^p \, dt \\
&=&
K^{\pmb{1}}\cdot \sum\nolimits_{k=0}^{K-1} \int\nolimits_{k/K}^{(k+1)/K} 
\int\nolimits_{k/K}^{(k+1)/K}  \big| f(s) - f(t) \big|^p\,ds \, dt \\
&\leq&
K \cdot 
\sum\nolimits_{k=0}^{K-1} \int\nolimits_{k/K}^{(k+1)/K} 
\int\nolimits_{-1/K}^{+1/K} \big| f(t+h \bmod 1) - f(t) \big|^p\,dh \: dt \\
&=&
K \cdot \int\nolimits_{-1/K}^{+1/K}
\big\| \tau_h f-f \big\|_p^p \, dh
\quad\leq\quad 2\sup_{|h|\leq1/K} \big\| \tau_h f-f \big\|_p^p 
\end{eqnarray*}
$\leq2^{-pn}$ for $K:=2^{\mu^{(p)}_f(n+1)}$.
\item[d)]
Estimate the $\ELL{p}$-modulus of a $K$-step function
$\displaystyle g_K=\sum\limits_{k=0}^{K-1} c_k \cdot\chi_{[k/K;(k+1)/K)}$:
\begin{eqnarray*}
\| g_K - \tau_\delta g_K \|^p_p
&=&
\sum\nolimits_{k=0}^{K-1} \delta \cdot\big|c_{k+1 \bmod K}-c_k\big|^p  \\
&\overset{\eqref{e:Hoelder0}}{\leq}& \delta 2^{p-1} \cdot \sum\nolimits_k \big(|c_{k+1 \bmod K}|^p+|c_k|^p\big) 
\;\;=\;\; K \delta 2^{p} \|g_K\|_p^p  
\end{eqnarray*}
$\leq 2^{2p} K \delta$
provided that $0<\delta\leq1/K$ and $\|g_K\|_p\leq2$.
Now $2^{2p} K \delta{\leq} 2^{-pn}$
for $\delta\leq 2^{-pn-2p-\Log K}$; 
hence $\mu^{(p)}_K(n)\leq pn+2p+\Log K$.
With $K:=2^{\sigma(n)}$,
Fact~\ref{f:Frechet}e) yields the claim.
\qed\end{enumerate}

\section{Perspectives}
\label{s:Perspectives}

We have explored the synergy between second-order complexity theory and classical approximation theorems (Arzel\`{a}-Ascoli and Fr\'{e}chet--Kolmogorov) 
in the context of \emph{moduli} for space of integrable functions. 
Specifically, parameterizing a separable (but not $\sigma$-compact) Banach space
via monotone compact covers naturally induces a \emph{second-order} notion of complexity, 
with runtime (or other resource) bounds depending on both the precision $n$ and said structural `modulus/rate' parameter.
Corollary~\ref{c:Equivalent} shows that
three different but natural such parameterizations 
($\ELL{p}$-modulus, rate of Fourier sequence convergence, rate of piecewise approximation)
are equivalent up to constant factors in both argument and value.
Consequently, each one yields the same notion of second-order linear (or, more coarsely, polynomial) time complexity.
That answers Question~\ref{q:Main} from the introduction
and enables the complexity theory of integrable functions.

\begin{remark}
\label{x:Compact}
We focus on function spaces over compact domains $X$ like $\IT$:
see also Subsection~\ref{ss:Dimension} below.
Here Arzel\`{a}-Ascoli and Fr\'{e}chet--Kolmogorov apply,
and yield moduli $\mu:\IN\nearrow\IN$.
A $\sigma$-compact domain like $\IR$
admits a countable monotone cover $X=\bigcup_k X_m$ by compact $X_m\subseteq X_{m+1}$:
cmp. Subsubsection~\ref{sss:SigmaCompact}.
Combining the family of moduli $\mu_m$ associated with each $X_m$
yields one joint bivariate modulus $(m,n)\mapsto \mu_m(n)$, 
i.e., of type $\IN\times\IN\nearrow\IN$.
Pre-composing the latter with the inverse of Cantor's pairing function
$\pi(n,m)=(n+m)\cdot(n+m+1)/2$ gives back type $\IN\nearrow\IN$.
\end{remark}

\subsection{Examples}
\label{ss:Examples}

According to Equation~\eqref{e:Embeddings},
the modulus of continuity is an upper bound
to the $\ELL{p}$-modulus. 
H\"{o}lder-continuous functions 
(Example~\ref{x:Modulus}c)
therefore also have `small' moduli/rates
when considered embedded in $\ELL{p}$.
Similarly to Example~\ref{x:Modulus}d),
the present subsection constructs integrable functions 
with `large' moduli/rates.
By Theorem~\ref{t:Equivalent}, a function with large $\ELL{p}$--modulus
also has large Step/Fourier rate, and vice versa.

\begin{example}
\label{x:Lacunary}
Consider 
\[
  f(t)\;:=\;\sum\nolimits_{k\geq0} \exp\bigl(2\pi i t\,2^k\bigr)/{2^k} \;\in\;\Lrm{2}(\IT,2) \enspace .
\]
The lacunary series forces its 
Hilbert Fourier rate
$\varphi_f^{(p)}(n)$ to grow exponentially in $n$. Similarly, 
\[ \sum\nolimits_{K\geq0} \exp\bigl(2\pi i t\,2^{2^k}\bigr)/{2^{2^k}} \;\in\;\Lrm{2}(\IT,2) \]
has double exponential Fourier rate.
\end{example}
Note that, like Liouville's historically first `explicit' construction of transcendental numbers,
our example is admittedly `artificial'.
The following construction, based on Example~\ref{x:Modulus}d+e), is arguably less artificial:

\begin{example}
\label{x:Exponential}
The derivative $h'$ on $(0;1]$ of $h(t)=1/\ln(e/t)$ from Example~\ref{x:Modulus}d)
belongs to $\ELL{1}(\IT)$, but not to $\calC(\IT)$.
By the \emph{Fundamental Theorem of Calculus},
its $\ELL{1}$-modulus is essentially 
the modulus of continuity of $h$,
and therefore also grows exponentially in $n$.

Similarly, $(h\circ h)'$ belongs to $\ELL{1}(\IT)$
and has doubly exponential $\ELL{1}$-modulus.
\end{example}
Our final Example~\ref{x:ONB} demonstrates that, over the Hilbert space $\ELLtwo$,
proceeding from the Fourier $\exp(2\pi ikt)$ orthonormal basis 
to a different orthonormal basis may drastically affect the asymptotic rate of approximation
in the sense of Definition~\ref{d:SchauderRate} below:

\begin{example}
\label{x:ONB}
Take any $f_0\in\ELLtwo$ with `large' (e.g. single or doubly exponential)
$\ELLtwo$-modulus $\mu^{(2)}_f$. Extend this single element $f_0$ to some orthonormal basis $f_k$
of $\ELLtwo$, $k\in\IN$.
\\
By Corollary~\ref{c:Equivalent}, $f_0$ has also `large' rate $\varphi^{(2)}_f$ of $\exp(2\pi ikt)$--approximation.
But in the orthonormal basis $f_K$, $f_0$ has coefficient sequence $(1,0,0,\ldots)$:
constant rate $\beta$ of $(f_k)$--approximation.
\end{example}
It remains to dually construct some $f\in\ELLtwo$ with $\ELLtwo$-modulus `small'
but asymptotically large rate $\beta$ of approximation w.r.t. some orthonormal basis of $\ELLtwo$.

\subsection{Parameterizing Sobolev and Banach Spaces}
\label{ss:Banach}

Definitions~\ref{d:LpModulus} and \ref{d:StepRate} and \ref{d:FourierRate}
and Corollary~\ref{c:Equivalent}
about $\ELL{p}$-moduli and Step/Fourier rates generalize from $\ELL{p}(\IT)$ 
to Sobolev spaces $\SOB{k,p}(\IT)$:
consider the $k$-th weak derivative of the functions under consideration;
cmp. \cite[\S4.1]{Ste17}.

In fact Definitions~\ref{d:StepRate} and \ref{d:FourierRate} 
as well as Subsubsection~\ref{sss:Uniform}
are instances of the following unifying general setting:

\begin{definition}
\label{d:SchauderRate}
Fix a Banach space $\calB$ and 
some sequence $B=(b_k)\subseteq\calB$ whose linear span is dense in $\calB$.
For $K\in\IN$ let $B_{K}\subseteq\calB$ denote the linear subspace 
spanned by $b_0,\ldots,b_{K-1}$. 

The (binary) \emph{rate of $B$--approximation} (``$B$--rate'')
$\beta_f$ of $f \in \calB$
is the pointwise least $\beta:\IN\nearrow\IN$ such that
there exists some $f_n\in B_{2^{\beta(n)}}$ 
with $\|f-f_n\|\leq2^{-n}$.
\end{definition}
Indeed, monomials form a Schauder basis to the Banach space $\calC[0;1]$ 
in Subsubsection~\ref{sss:Uniform}; and the 
characteristic functions of intervals $\big[k/K;(k+1)/K\big)$, 
although not linearly independent as both $k$ and $K$ vary, 
do span a dense subspace of Banach space $\ELL{p}(\IT)$.

Example~\ref{x:ONB} below demonstrates that 
the quantitative equivalence among moduli/rates
according to Corollary~\ref{c:Equivalent} 
is particular to certain Schauder bases of $\ELL{p}$.

\begin{question}
\label{q:General}
Which further `natural' Schauder bases of $\ELL{p}$ 
induce (binary) rates linearly equivalent to the above ones?
\end{question}

\subsection{Higher Dimensions}
\label{ss:Dimension}

Section~\ref{s:Integrable} considers $\ELL{p}(X)$ for $1<p<\infty$ 
with $X=\IT$ the real or $X=\IS$ the complex unit circle: one-dimensional domains.
Definitions~\ref{d:LpModulus} and \ref{d:StepRate} and \ref{d:FourierRate}
extend straightforwardly to the $d$-dimensional torus (=hypercube with `wraparound'):

\begin{definition}
\label{d:Dimension}
\begin{enumerate}
\item[a)]
Let $\IT^d=\prod^d\IT=[0;1)^d\bmod\vec 1$ denote the $d$-dimensional torus,
and $\big(\tau_{\vec\delta} f\big)(\vec x) = f(\vec x+\vec\delta \bmod \vec 1)$
the shift by vector $\vec\delta$ with wraparound.

The (binary) $\Lrm{p}$-modulus $\mu=\mu^{(p)}_f:\IN \nearrow \IN$ 
of $f \in \Lrm{p}(\IT^d)$ is pointwise minimal such that
$\|f-\tau_{\vec\delta} f\big\|_p\leq2^{-n}$ for all $|\vec\delta|\leq2^{-\mu(n)}$.
\item[b)]
For $K\in\IN$ let $\calS_{K}(\IT^d)$ denote the vector space of complex functions
which are constant on each of the $K^d$ half-open sub-cubes 
$\big[\vec k/K;(\vec k+\vec 1)/K\big)\subseteq\IT^d$, $0\leq k_1,\ldots,k_d<K$.

The (binary) rate of piecewise approximation (``Step rate'')
$\sigma^{(p)}_f$ of $f \in \Lrm{p}(\IT^d)$
is the pointwise least $\sigma:\IN\nearrow\IN$ such that
there exists some $f_n\in\calS_{2^{\sigma(n)}}(\IT^d)$ 
with $\|f-f_n\|_p\leq2^{-n}$.
\item[c)]
For $f\in\ELL{1}(\IT^d)$ let
\[ 
\hat f_{\vec k} \;\;=\;\; \int\nolimits_{\vec 0}^{\vec 1} f\big(\vec x\big)\cdot\exp\big(-2\pi i \langle\vec k,\vec x\rangle\big)\,d\vec x 
\;\;\in\;\IC
\]
denote the $\vec k$-th Fourier coefficient, $\vec k\in\IZ^d$. And write
\begin{equation}
\label{e:Fourier2}
\calF_K f(\vec x) \;\;=\;\; \sum\nolimits_{|\vec k|\leq K} \hat f_{\vec k}\cdot \exp\big(2\pi i \langle\vec k,\vec x\rangle\big) 
\end{equation}
for the $K$-th partial Fourier sum.

The (binary) rate of Fourier series convergence (``Fourier rate'')
$\varphi_f^{(p)}$ of $f\in\ELL{p}(\IT^d)$
is the pointwise least $\varphi:\IN\nearrow\IN$ 
with $\displaystyle\big\|f-\calF_{2^{\varphi(n)}} f\big\|_p\leq2^{-n}$.
\end{enumerate}
\end{definition}
Note the ambiguities in (a) and (c) regarding the particular norms in $|\vec\delta|$ and $|\vec k|$.
Indeed, although all norms in dimension $d$ are equivalent, 
it is known \cite{Fefferman71} that from $d=2$ on
the Fourier sums $\lim_{K\to\infty}\calF_K f$ from Equation~\eqref{e:Fourier2} 
may fail to converge in $p$-norm for $f\in\ELL{p}(\IT^d)$ when $p\neq2$ 
in case $|\vec k|=\sqrt{k_1^2+k_2^2}$ denotes the Euclidean norm.
They do converge in any dimension $d$ (when $1<p<\infty$ like in the case $d=1$)
in case $|\vec k|=\max\{|k_1|,\ldots,|k_d|\}$ is understood to denote the maximum norm---which also explicitly underlies (b).
With this convention, Corollary~\ref{c:Equivalent} generalizes from $\IT$ to $\IT^d$
for any fixed $d$, where now the constants hidden in $\calO()$ depend on both $p$ and $d$.

\begin{remark}
\label{r:Shapes}
Extending merely Definitions~\ref{d:LpModulus} and \ref{d:StepRate} and \ref{d:FourierRate},
not to mention Theorem~\ref{t:Equivalent}, from cubes/tori $\IT^d$ to other geometric shapes
becomes challenging already for, say, planar triangles $X$ \cite{Babb23}
and exceeds the scope of this work.
\end{remark}


\addcontentsline{toc}{section}{Bibliography}
\bibliographystyle{my_alpha}
\bibliography{cca,bib_aras,bib_martin}

\end{document}